# Influence of oxygen ion implantation on magnetic microstructure in Pt/Co/Pt multilayers with perpendicular magnetic anisotropy

Anmol Sharma[a], Mukul Gupta[b], Prasanta Karmakar[c], V. Raghavendra Reddy[b], Vivek K. Malik[d], Andrei Gloskovskii[e], Ranjeet Kumar Brajpuriya[a], Ajay Gupta[a], Vishakha Kaushik[f], and Sachin Pathak[a]*

[a]Department of Physics, UPES Dehradun, Uttarakhand, 248007, India
[b]UGC-DAE Consortium for Scientific Research, Indore, 452017, India
[c]Ion Beam Development and Application Section, RIBF Group, Variable Energy Cyclotron Centre, HBNI, 1/AF, Bidhannagar, Kolkata 700064, India
[d]Department of Physics, Indian Institute of Technology Roorkee, Roorkee, 247667, India
[e]Deutsches Elektronen-Synchrotron DESY, Hamburg, 22607, Germany
[f]Materials & Nano Engineering Research Laboratory, Department of Physics, School of Physical Sciences, DIT University, Dehradun- 248009, India

*Corresponding author: s.pathak@ddn.upes.ac.in (Sachin Pathak)

**Abstract:** The interaction of oxygen with cobalt and cobalt-based alloys has been a very important topic in the field of spintronics as it leads to enhanced orbital anisotropy and interfacial Dzyaloshinskii-Moriya interaction (DMI), which are crucial in the context of applications such as magnetic tunnel junctions (MTJs) based data storage and domain wall (DW) motion. To understand the complex and interesting relationship between oxygen and ferromagnetic (FM)/heavy metal (HM) interfaces, we studied controlled oxygen ion ($O^+$) implantation in a cobalt layer located in a Pt/Co 1.2 /Pt (nm) multilayer with a specific structure. At high implantation fluence, the perpendicular anisotropy was lost, as verified by in-plane hysteresis measurements. Under low magnetic field conditions, the DW dynamics of Co/Pt multilayers were analyzed, highlighting key parameters such as DW velocity, roughness amplitude, and roughness exponent. After $O^+$-ion implantation, the DW velocity increased by more than 50 times, rising from 5 µm/s to 300 µm/s compared with the as-deposited multilayer. The fundamental cause of this improvement is the structural and magnetic changes brought by the implantation, which successfully lower the energy barriers preventing DW movements. The results show how oxygen implantation can be used to precisely tailor the ferromagnetic interfaces, leading to promised improvements in the functionality of next-generation spintronic devices.

**Highlight of research:**

- The effect of oxygen ion implantation at 6 keV energy in Pt/Co/Pt multilayer with perpendicular magnetic anisotropy (PMA) was studied.
- For low-fluence oxygen implantation, the coercivity of the multilayer was reduced with retained PMA.
- For high-fluence oxygen implantation, anisotropy switches from perpendicular to in-plane.
- DW dynamics were investigated in the creep regime and in the PMA retained Co/Pt$_{low}$ multilayer. DW velocity was enhanced from 5 $\mu$m/s to 270 $\mu$m/s compared to the pristine multilayer.
- Roughness analysis shows exponent $\zeta \approx 0.70$ with higher amplitude after implantation, indicating modified pinning landscapes that enable faster but rougher DW motion.

**Graphical abstract**

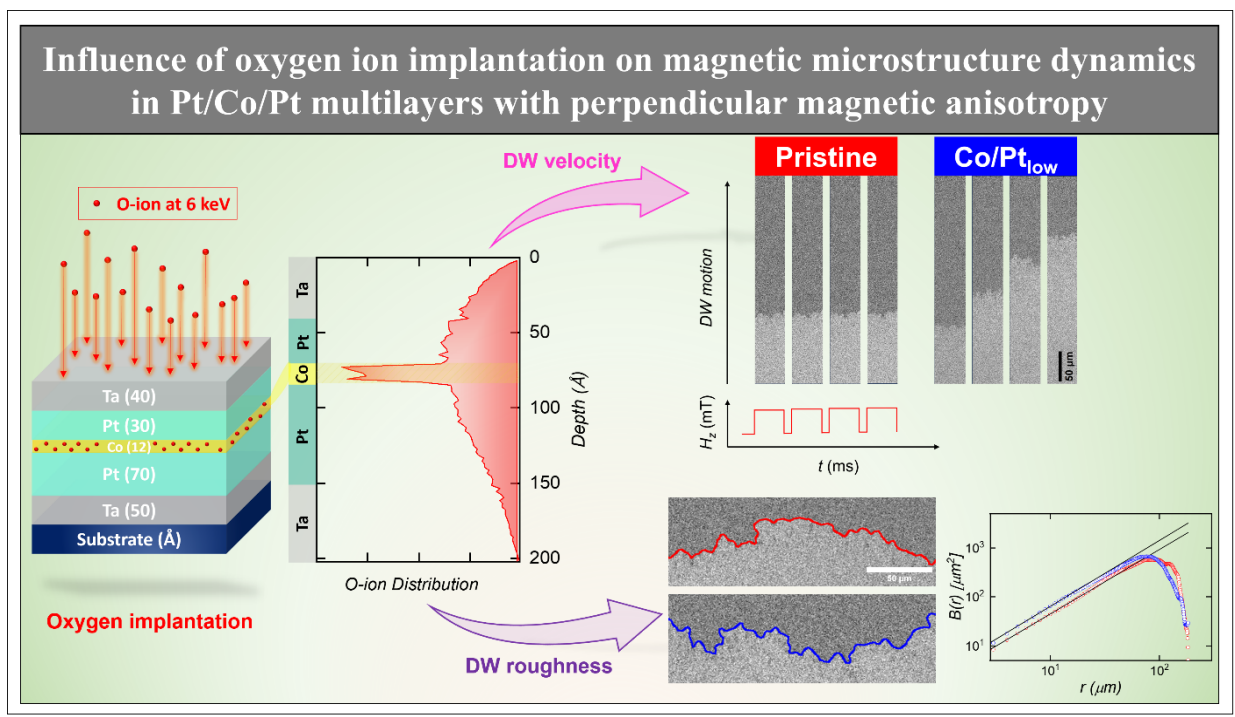

1. **Introduction**

Advancements in material science for spintronic technologies demand materials with enhanced functionalities compared to those used in earlier generations. Multilayers consisting of FM layers interfaced with HMs have emerged as particularly promising candidates for next-generation spintronic applications, including high-density data storage and magnetic sensors. Their appeal lies in the tunable structural and magnetic properties that arise from interfacial interactions between FM and HM layers[1–3]. Systems with tunable and high anisotropy are considered best for such applications, and there comes perpendicular magnetic anisotropy (PMA), which is considered to be influenced by underlayer phase orientation, such as in the case of Pt and Pd, fcc (111) is considered to have strong PMA in these multilayers[4,5]. In addition to PMA, other interfacial phenomena, such as the DMI, further affect the dynamics of nanoscale magnetic textures, including DWs and skyrmions, which are vital for spintronic device applications.

One effective strategy for tailoring such interfacial phenomena is to engineer the chemical and structural environment of the FM and HM layers. Several approaches have been explored, including impurity introduction during deposition[6], insertion of alternative HM layers[7,8] (e.g., W, Ta, Pd, or Pt) to break structural symmetry, and post-deposition treatments[9]. In particular, modification during deposition has been widely employed, for instance, by adding Cu impurities during Co deposition to tune anisotropy or by introducing oxygen after FM deposition[10,11]. Similar approaches have also been applied to HM layers, where controlled doping or gas exposure alters their structural and magnetic properties[12,13]. Beyond deposition-based techniques, voltage-driven magneto-ionic transport has also shown promise for dynamically tuning magnetic properties[14]. Collectively, these methods highlight the importance of interfacial engineering in optimizing PMA, DMI, and spin-orbit torque (SOT) for spintronic-based device applications.

The choice of impurity and the method of its introduction are critical in determining the effectiveness of such modifications. Recent studies have demonstrated that oxygen plays a particularly significant role in tuning the interfacial properties of both FM and HM layers[15,16]. For example, controlled oxidation of cobalt and cobalt-based alloys has been shown to enhance PMA and DMI, while oxygen modification of Pt underlayers has been linked to improved SOT efficiency[6,17]. Such findings suggest that oxygen incorporation at well-defined interfaces can provide a powerful route for tailoring the structural and magnetic behaviour of multilayer systems. However, achieving precise control over oxygen incorporation remains a challenge, as conventional deposition techniques may lead to non-uniform distribution and unintended structural effects.

Based on the importance of previous works, we have selected oxygen as an impurity within the multilayer, and post-deposition impurity added using the implantation technique rather than depositing FM+O or HM+O. This method enables controlled modification of the Co interface without altering the overall deposition process. We systematically investigate the impact of oxygen implantation on the structural and magnetic properties of Pt/Co/Pt multilayer, with particular emphasis on PMA and DW dynamics, including velocity and roughness. The results are compared with the pristine Pt/Co/Pt reference multilayer to reveal the role of interfacial oxidation in shaping the magnetic behaviour. Our findings provide new insights into the controlled tuning of interfacial phenomena, offering a promising pathway toward the design of optimized magnetic multilayers for future spintronic applications.

## 2. Experimental details

Multilayer with PMA was deposited on $Si/SiO_2$ substrate at room temperature using ultrahigh-vacuum DC magnetron sputtering with a base pressure of $\approx 5 \times 10^{-8}$ torr. The multilayer stack comprised Ta as a buffer and capping layer, with the final structure described as $Si/SiO_2$/Ta (50)/Pt (70)/Co (12)/Pt (30)/Ta (40 Å). Prior to deposition of the multilayer, each target was calibrated to achieve the mentioned thickness. The sputtering power was set to 60, 30, and 20 W for Ta, Pt, and Co, respectively, under a constant argon (Ar) gas flow of 50 sccm while maintaining a working pressure of $\approx 2.5 \times 10^{-3}$ torr, which resulted in deposition rates of 0.294, 0.250, and 0.125 Å/sec for each layer in sequence. To ensure the uniformity of the layers, during deposition, the substrate was rotated at a rpm of 20. After deposition of the multilayer, oxygen ions ($O^+$) with an energy of 6 keV were implanted into the films using a mass-analyzed $O^+$ beam generated by a 2.4 GHz electron cyclotron resonance (ECR) plasma source. The ions were incident normal to the film surface, and the irradiation was carried out under high-vacuum conditions with a pressure of approximately $6 \times 10^{-7}$ mbar.

The structural characteristics of the pristine as well as implanted multilayer were evaluated using the grazing incidence X-ray diffraction (GIXRD) technique. The instrument (Panalytical, Intelligent diffractometer, Model: Emp3, Netherlands) with Cu-$K_\alpha$ radiation at 1.5405 Å was equipped with a 1Der detector. Along with this, the instrument is equipped with a multilayer mirror to produce a parallel beam and a parallel plate collimator on the detector side to ensure a monochromatic Cu-$K_\alpha$ radiation beam. For measuring the hysteresis loop as well as the DW analysis, a Magneto-optical Kerr effect (MOKE) microscope system (M/s Evico Magnetics, Germany) was utilised, which is accessible at UGC-DAE CSR Indore[18]. The device consists of a digital CCD camera (Hamamatsu ORCA-03G) and employs a white LED light source.

To confirm the controlled $O^+$-ion implantation along with alloy formation at the interfaces due to intermixing, HAXPES was performed at the P22 beamline of the PETRA III synchrotron in Germany. A double-crystal monochromator (DCM) employing a Si (111) crystal produced a highly monochromatic X-ray beam with an energy of 6 keV. To ensure an uncontaminated atmosphere for measurements, the analysis chamber was kept at an ultra-high vacuum of greater than $5 \times 10^{-9}$ torr. Considering the low thickness of the Co layer, HAXPES spectra were taken with 20 sweeps at the Co-edge to achieve high-statistical data. The energy resolution of this beam was $\Delta E = 100$ meV. Data analysis was done using CasaXPS[19] Version 2.3.26PR1.0, with a Shirley background subtraction and a Lorentzian Asymmetric lineshape applied for accurate peak fitting.

## 3. Result and Discussion

For the implantation of oxygen within the Co/Pt multilayer, the energy and fluence were selected on the basis of computational analysis, which was performed using the transport of ions in matter (TRIM) code[20] shown in Figure 1(a). The chosen implantation energy was optimized to provide a homogeneous distribution of $O^+$ ions within the Co layer, ensuring that the implantation effects at the upper and lower Pt/Co interfaces remain consistent. To evaluate the influence of $O^+$ implantation on the magnetic and structural properties of the multilayer, the ion fluences were chosen based on an estimate of the total number of Co atoms in the FM layer, which is on the order of $10^{16}$ atoms for a $1 \times 1$ cm$^2$ sample. TRIM simulations indicate that, for a given fluence, approximately 25% of the incident $O^+$ ions are implanted within the Co layer. On this basis, fluences of $\approx 8.72 \times 10^{14}$ and $\approx 2.18 \times 10^{15}$ ions/cm² were selected to represent low and high implantation levels, and the corresponding multilayers are denoted as Co/Pt$_{low}$ and Co/Pt$_{high}$ in the subsequent analysis.

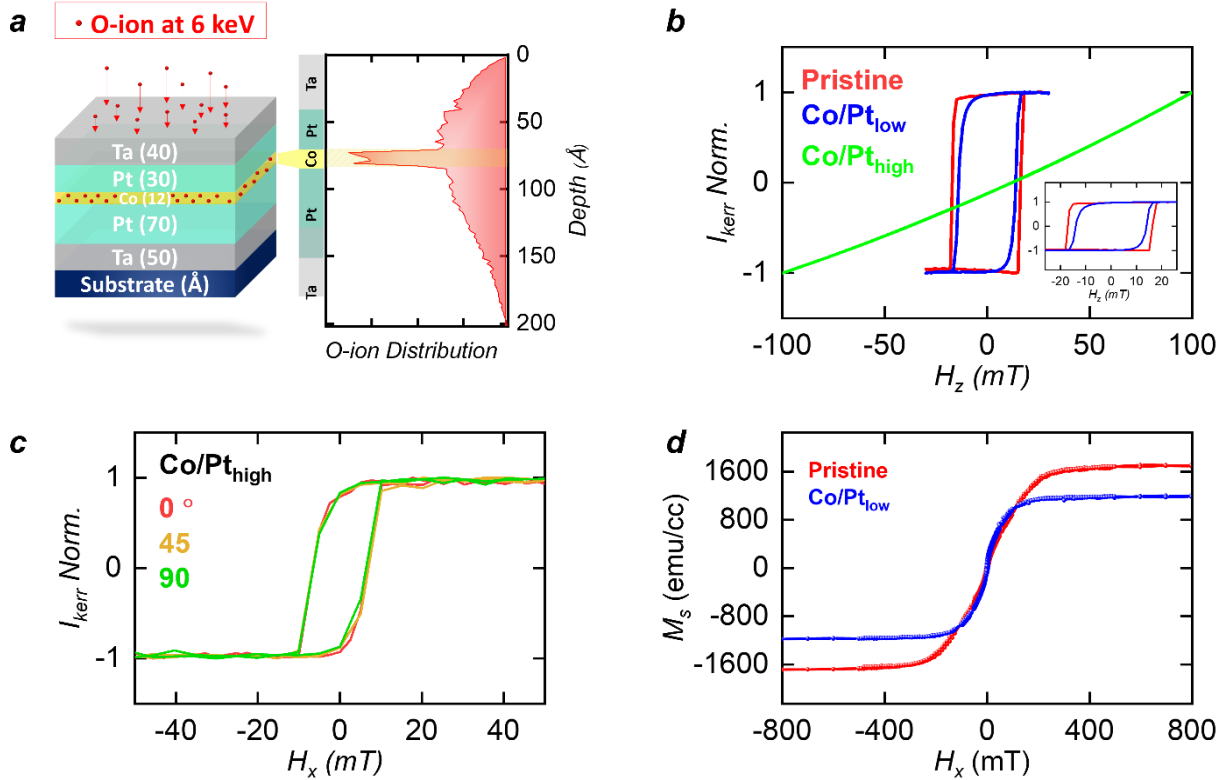

**Figure 1: (a)** Schematic of Co/Pt multilayers implanted with SRIM-simulated depth profile of 6 keV O$^+$ ion implantation, highlighting maximum ion accumulation (yellow) within the Co layer. **(b)** Comparison of PMOKE hysteresis loops for the pristine (red) and O-ion implanted Co/Pt$_{low}$ (blue) and Co/Pt$_{low}$ (green) multilayer. **(c)** Longitudinal MOKE (LMOKE) hysteresis loops of the Co/Pt$_{high}$ multilayer measured under an in-plane magnetic field applied at 0°, 45°, and 90° **(d).** In-plane SQUID-VSM hysteresis loops of the pristine and Co/Pt$_{low}$ multilayer.

Figure 1(b) illustrates the hysteresis loops measured in perpendicular geometry. The pristine multilayer, which exhibited clear PMA after deposition, shows a square hysteresis loop, and the Co/Pt$_{low}$ multilayer also retains a square loop after implantation, confirming the preservation of PMA. The coercivity decreased from 17 mT in the pristine sample to 14 mT in the Co/Pt$_{low}$ multilayer, which suggests that low-fluence treatment reduces the nucleation barrier for magnetization reversal, thereby slightly softening the magnetic switching process. However, at higher fluence levels, no hysteresis loop was observed in perpendicular geometry, indicating a complete loss of PMA. To confirm whether the loss of PMA was due to overoxidation, which leads to degradation of ferromagnetism or a reorientation of the easy axis of magnetization from perpendicular to the in-plane axis of the sample, additional measurements were performed with an in-plane magnetic field. The longitudinal MOKE (LMOKE) loops of the Co/Pt$_{high}$ multilayer at three distinct angles, 0, 45, and 90º, are displayed in Figure 1(c). The reorientation of magnetization into the film plane with isotropic anisotropy is confirmed by a distinct hysteresis loop with a coercivity of almost 6 mT under in-plane geometry. Table 1 summarises the magnetic parameters obtained from MOKE measurements for all the multilayers.

**Table 1:** Summary of multilayer codes, implanted oxygen fluences (in ion/cm$^2$), and the magnetic parameters extracted from PMOKE and LMOKE measurements.

| Sr. No. | Multilayer code | Fluence (ion/cm$^2$) | Coercivity (mT) | $M_r/M_s$ | Anisotropy |
|---|---|---|---|---|---|
| 1. | Pristine | - | 17.0 | 1.00 | PMA |
| 2. | Co/Pt$_{low}$ | $8.72 \times 10^{14}$ | 14.0 | 1.00 | PMA |
| 3. | Co/Pt$_{high}$ | $2.18 \times 10^{15}$ | 5.7 | 0.74 | IMA |

To further study the influence of oxygen implantation on the magnetic parameters, Figure 1(d) presents the in-plane SQUID-VSM measurements for the pristine and Co/Pt$_{low}$ multilayers. The pristine film exhibits a saturation magnetization ($M_s$) of 1690 emu/cc and a saturation field of 280 mT, characteristic of strong PMA in Co/Pt systems. In comparison, the Co/Pt$_{low}$ oxygen-implanted sample shows reduced values of 1180 emu/cc and 180 mT, reflecting a partial weakening of interfacial anisotropy and magnetic moment. These trends are consistent with the behaviour observed in Figures 1(b) and 1(c), collectively demonstrating that low-fluence oxygen implantation reduced but does not eliminate the PMA in the multilayer.

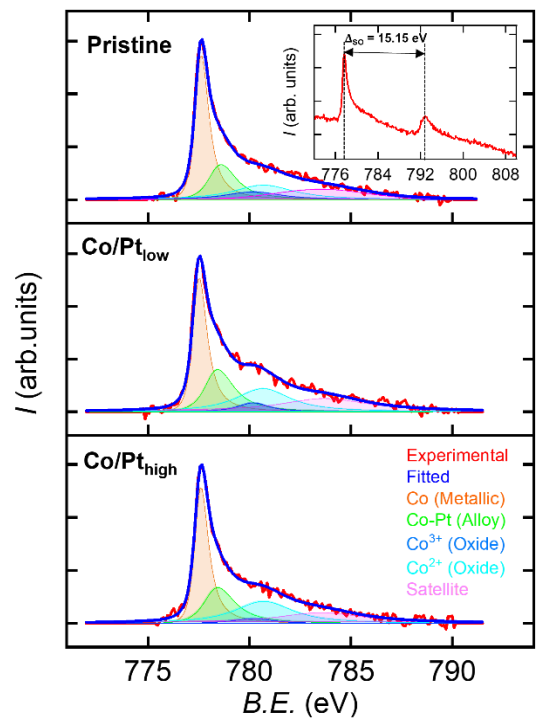

**Figure 2:** Experimental and best-fit Co-2$p_{3/2}$ HAXPES spectra acquired at 6 keV photon energy for the pristine and O$^+$-ion implanted multilayers at low and high fluences.

For further confirmation that the loss of PMA at high fluence was not caused by overoxidation, HAXPES measurements were carried out. This chemical investigation complements the magnetic hysteresis results, where in-plane measurements (Figure 1c) already indicated that the disappearance of the out-of-plane hysteresis loop originated from a reorientation of the magnetic easy axis rather than from oxidation. To directly probe possible implantation effects, the photon energy in HAXPES was chosen by considering the inelastic mean free path (IMFP) of photoelectrons in each layer. At an excitation energy of 6 keV, the IMFPs for Co, Pt, and Ta were approximately 60.4 Å, 52.7 Å, and 60.2 Å, respectively, ensuring sufficient sensitivity to detect chemical modifications throughout the multilayer[21].

Figure 2 presents the experimental Co-$2p_{3/2}$ spectra along with their fitted data for both pristine and $O^+$ ion-implanted multilayers. The Co-$2p_{3/2}$ HAXPES spectra exhibit prominent components corresponding to the $2p_{3/2}$ and $2p_{1/2}$ core levels, located at binding energies (B.E.) of approximately 777.72 eV and 792.87 eV, respectively (as shown in the inset of Figure 2). The energy separation between these two components gives a spin-orbit splitting ($\Delta_{SO}$) of about 15.15 eV. The fitting procedure included four main contributions: metallic Co at ~777.6 eV, two oxide components corresponding to $Co^{2+}$ and $Co^{3+}$ around ~780.7 and ~780.2 eV, a Co-Pt alloy signal, and a satellite near 783.5 eV, attributed to charge-transfer processes between Co and oxygen[22], which become more pronounced upon $O^+$ ion implantation. A slight increase in intensity of oxide peaks was observed in the implanted samples, suggesting minor oxygen presence at the interfaces. However, keeping the overall chemical environment unchanged suggests that no overoxidation occurred. Table 2 summarizes the fitting parameters, which confirm that oxygen implantation is restricted to minor interfacial modifications. Therefore, combining the magnetic hysteresis loops and HAXPES measurements, these results conclude that the change in anisotropy of the multilayer at high fluence originates from interfacial chemical and structural modifications.

**Table 2:** Fitting parameters for the Co-$2p_{3/2}$ HAXPES spectra of the sample in both its pristine and $O^+$-ion implanted multilayer.

| Sr. No. | Multilayer | Co % conc. | Co-Pt alloy % conc. | $Co^{3+}$ % conc. | $Co^{2+}$ % conc. |
|---|---|---|---|---|---|
| 1. | Pristine | 39 | 19 | 8 | 19 |
| 2. | Co/Pt$_{low}$ | 33 | 19 | 5 | 23 |
| 3. | Co/Pt$_{high}$ | 34 | 20 | 4 | 26 |

To investigate the structural characteristics of the Co/Pt multilayers, GIXRD measurements were carried out at a fixed incidence angle of 0.5° for all samples. The GIXRD measurements for the pristine multilayer and the two $O^+$ ion-implanted multilayers, labelled as Co/Pt$_{low}$ and Co/Pt$_{high}$, are shown in Figure 3. It is well established that PMA in Co/HM (HMs=Pt or Pd) systems is supported by a fcc structure with a (111) preferential orientation in both Co and HM layers[23]. However, due to the extremely thin individual layer thicknesses, diffraction peaks from Co and Pt often overlap, making them difficult to resolve separately in multilayer stacks. Two prominent peaks can be observed in the measured GIXRD patterns, located near 37° and 67°. The peak around 67° aligns with the Pt (220) reflection, consistent with earlier reports[24]. The peak around 37° is likely due to overlapping contributions from Ta (002)/(110) and Pt and Co (111) orientations, making it difficult to identify clearly. Crucially, the overall diffraction pattern is mostly unaffected even after high $O^+$ ion implantation. This implies that the loss of PMA in Co/Pt$_{high}$ multilayer does not involve major alterations to the diffraction pattern, suggesting that PMA degradation is likely due to interfacial modifications.

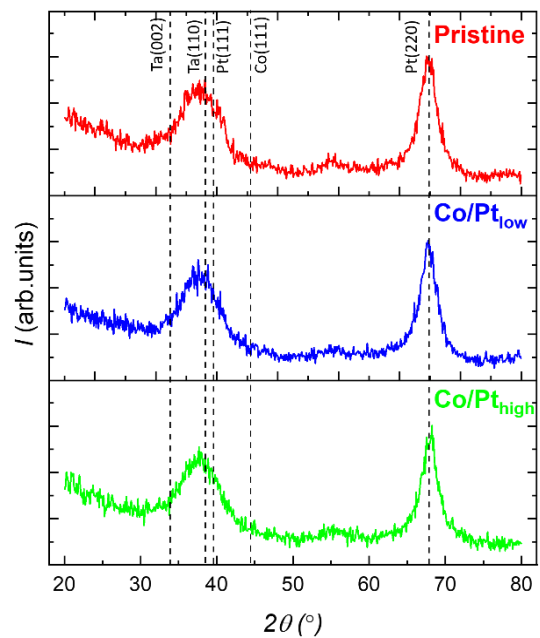

**Figure 3:** GIXRD patterns of pristine and O-ion implanted Co/Pt multilayers at low and high fluences. The black dashed lines mark the reference peak positions corresponding to the expected crystalline phases.

Since magnetic microstructures have gained a lot of attention in recent years, DW dynamics were also studied as a function of $O^+$ ion implantation using PMOKE microscopy. As it has been discussed above, for only low implantation of $O^+$ ion in the multilayer, PMA remains intact, we have analyzed these DW dynamics for pristine and Co/Pt$_{low}$ multilayer. DWs' dynamics in these multilayers are influenced by different means, like external magnetic field, defects, temperature, and so on. For very low magnetic fields at finite temperature, these domains have to overcome some energy barriers to propagate and hence for their magnetization reversals. Therefore, at low driving field at finite temperature, the dynamics of DW velocity ($v$) can be determined using an Arrhenius law mathematically[25] given by:

$$v = v_d exp\left[-\frac{T_d}{T}\left(\left(\frac{H}{H_d}\right)^{-\mu} - 1\right)\right] \quad (1)$$

where, $v_d$ is the characteristic velocity, $T$ is the temperature, $H_d$ is the depinning field, $T_d$ is the depinning temperature, and $\mu$ is the universal creep constant. Taking $ln$ on both sides of equation (1) will give us:

$$ln(v) = ln(v_o) - \alpha H_z^{-\mu} \quad (2)$$

where, $ln\, v_o = ln\, v_d + T_d/T$ and $\alpha = T_d\, H_d^{\mu}/T$. It can be noticed that the above equation (2) is analogous to the equation of a straight line, $\alpha$ is the slope called creep scaling constant, and $ln(v_o)$ Is the intercept called depinning velocity? Thus, below the depinning field $H_d$, DW velocity data should follow the above equation (2) and are said to be in the creep regime. DW velocity in these magnetic multilayers is governed by thermal activation over disorder-induced energy barriers, characterized by an energy scale $k_B T_d$, where $k_B$ is the Boltzmann constant, and $T_d$ is the depinning temperature. The universal creep exponent $\mu = 1/4$ arises when the equilibrium roughness exponent $\zeta_{eq}$ equals 2/3. The value is commonly used by different research groups to model DWs in FM multilayers as elastic interfaces propagating through a disordered medium. Therefore, even below the $H_d$, i.e., when the external field is very small, we can still extract valuable information about the material properties and external parameters that influence DW dynamics, as well as validate the predictions of creep theory.

Figure 4 (a) shows the velocity curve for pristine and Co/Pt$_{low}$ multilayer, and it was observed that the motion was still in the creep regime (as shown in Figure 4 (a) inset). It was observed that, when a pristine multilayer was compared with a Co/Pt$_{low}$ multilayer, O$^+$ ion implantation leads to an increase in the DW velocity of the Co/Pt$_{low}$ multilayer, as shown by the yellow highlighted region. For example, under a fixed field of $H_z$ = 13.93 mT and a pulse duration of 0.2 s, the pristine sample showed a velocity of nearly 5 $\mu$m/s (shown in Figure 4 (b)). In contrast, after implantation, the velocity increased dramatically to nearly 300 $\mu$m/s. The DW velocity increased by more than 50 times, highlighting the significant role of ion implantation in modifying the interfacial properties and lowering the energy barriers for DW motion, thereby promoting faster DW propagation.

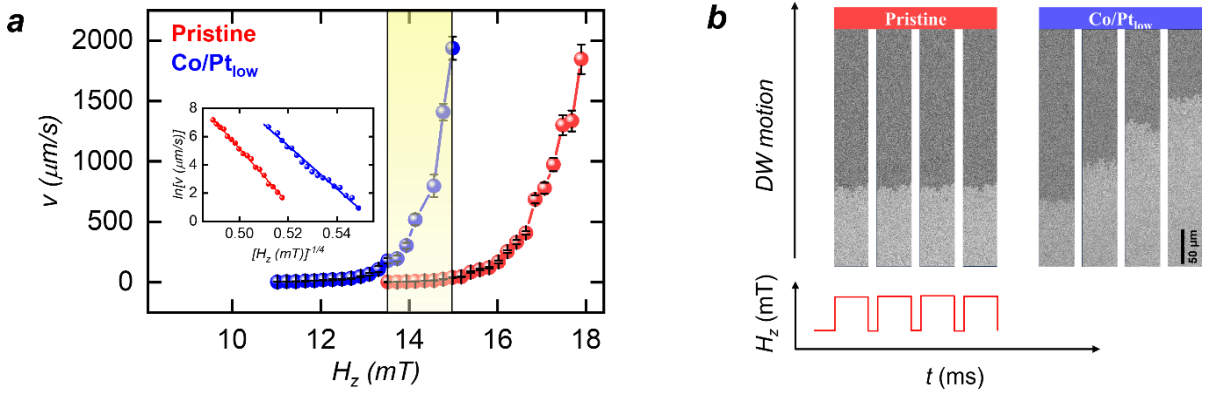

**Figure 4: (a)** DW velocity (*v*) variation as a function of applied $H_z$ for pristine and Co/Pt$_{low}$ multilayer (inset: plot of *ln(v$_o$)* versus $H_z^{-1/4}$ validating the creep law) (Yellow highlighted region shows velocity comparison between two multilayers) **(b)** DW velocity comparison between pristine and Co/Pt$_{low}$ multilayer at fixed $H_z$ =13.93 mT and pulse 0.2 s.

To correlate the velocity changes induced by O$^+$ ion implantation, we analyzed the DW dynamics using the height-height correlation function, [26], also referred to as the roughness function *B(r)*:

$$B(r) = \frac{1}{M-k} \sum_{i=1}^{M-k} [u(x_i + r) - u(x_i)]^2 \quad (4)$$

where $k = r/\delta < M$ is an integer value. For this analysis, a DW of length 185 μm was considered and discretized into equally spaced points $x_i = i\delta$, with $i = 1, 2, …, M$. The spacing between successive points was taken as $\delta = 1$ μm. The DW position at each point $x_i$, under a fixed $H_z$ field and pulse width, was denoted by $u(x_i)$. For self-affine DWs, for the low-*r* region, the roughness function is given by the power square law with the roughness exponent (*ζ*). Mathematically, the roughness function is expected to follow the scaling relation:

$$B(r) = B_o \left(\frac{r}{l_o}\right)^{2\zeta} \quad (5)$$

where a normalization length $l_o$=1 μm is introduced to ensure that both the roughness function *B(r)* and the roughness amplitude $B_o$ can be expressed in the same units. Figure 5(a) shows the DW profiles for pristine and Co/Pt$_{low}$ multilayer, which was used for toughness parameter analysis. For both pristine and Co/Pt$_{low}$ multilayer, the DW was expanded over five successive pulses, and five distinct profiles were selected for analysis. The roughness function had been calculated separately for each DW profile. These results were then averaged over the five profiles to obtain an average roughness function, which was subsequently used to determine the roughness parameters, i.e., the roughness exponent ($\zeta$) and amplitude ($B_o$). Figure 5(b) shows the average roughness functions for pristine (red curve) and Co/Pt$_{low}$ multilayers (blue curve), fitted using Eq. (5), where the slope of the curve gives the $\zeta$ value, and the intercept gives the $B_o$ value. For fitting, we have considered two methods, (i) considering the resolution of the microscope, i.e., 1.49 pixels/$\mu$m, in which each pixel represents approximately 0.671 $\mu$m, we have fixed the lower limit to the fifth pixel of the microscope and kept the upper limit free. While in (ii), since the power law gives a reliable analysis at low scaling length, we have manually selected a point near 14 $\mu$m and set the lower bound free. Thus, from the averaged roughness curves extracted from 5 DWs and combining both these fitting analyses gives us statistically good and reliable results, with the coefficient of determination ($R^2$) for the goodness of fit maintained above 0.998 for both the multilayers. It can be seen in Figure 5(a) that the DWs in the Co/Pt$_{low}$ multilayer appear rougher than those in the pristine multilayers. It was observed that both exhibit similar roughness exponents ($\zeta \approx 0.70$), but the roughness amplitude increases significantly from 1.83 to 2.32 $\mu$m$^2$ after implantation. The extracted values are summarized in Table 2.

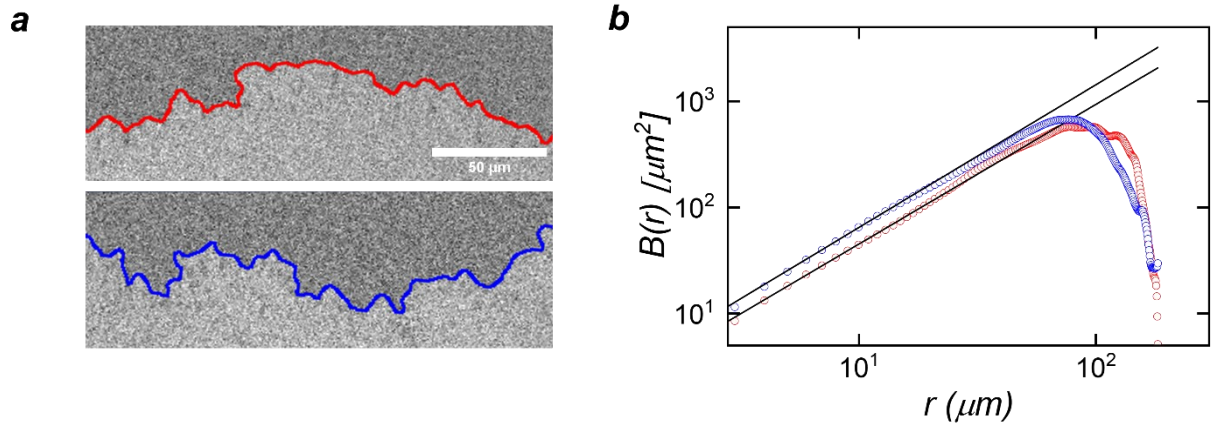

**Figure 5:** **(a)** DW profiles under an out-of-plane field $H_z$ (red: pristine, blue: Co/Pt$_{low}$ with a 50 $\mu$m scale bar) **(b)** Corresponding average roughness function $B(r)$ in log-log scale, from which the roughness exponent ($\zeta$) and amplitude ($B_o$) are extracted using linear fits with the coefficient of determination above $R^2 = 0.998$.

**Table 2:** Extracted roughness amplitude ($B_o$ in $\mu$m$^2$) and roughness exponent ($\zeta$) values obtained from power-law fitting of the pristine and Co/Pt$_{low}$ multilayer.

| Sr. No. | Multilayer | $B_o$ ($\mu$m$^2$) | $\zeta$ |
|---|---|---|---|
| 1. | Pristine | 1.83 | 0.70 |
| 2. | Co/Pt$_{low}$ | 2.32 | 0.72 |

Thus, combined with the earlier observations of reduced coercivity and enhanced DW velocity observed in Co/Pt$_{low}$ multilayers when compared with pristine Co/Pt multilayer, these results highlight a consistent trend, that is, after oxygen implantation in the Co/Pt multilayer, energy barriers for DW motion were reduced. The slight increase in roughness amplitude indicates disorders caused by O$^+$ ion implantation within the multilayer, the reduced coercivity and anisotropy lead to the higher DW velocity, suggesting that the effective pinning barriers are weakened, which enabled faster DW propagation. This behaviour can be explained by a redistribution of pinning sites induced by ion implantation, where implantation causes additional nanoscale disorder within the multilayers, increasing DW fluctuations while simultaneously decreasing the strength of individual pinning centers.

4. **Conclusion**

To conclude, we have studied a Pt/Co/Pt multilayer with O$^+$ ion implantation at 6 keV energy at two different fluences. It was observed that at low fluences, the multilayer retains its PMA properties with a reduction in its coercivity and perpendicular anisotropy. Further, FM domain dynamics were investigated in the as-deposited and PMA retained multilayer. It was observed that, due to a reduction in coercivity value, nucleation field for domain formation was reduced, which leads to an increase in velocity of domains at $H_z$ = 13.93 mT, from 5 $\mu$m/s to 270 $\mu$m/s when compared with the as-deposited multilayer. It was also observed that, after implantation, nucleation sites were also increased, which is commonly observed when such systems are irradiated. From DW roughness analysis, it was observed that, in DW, roughness was increased while roughness exponent remains similar, $\zeta \approx 0.70$ for both the as-deposited and Co/Pt$_{low}$ multilayer, indicating modified pinning landscapes which enable faster but rougher DW motion. Thus, the study highlights the role of oxygen interactions at the FM/HM interface in tuning magnetic anisotropy and domain dynamics, which is of growing interest for spintronic applications such as racetrack memory and DW-based logic devices.

**Acknowledgments**

AS is a Junior Research Fellow (JRF) supported by UPES, Dehradun. The authors would like to thank Rakhul Raj for helping to carry out DW measurements as well as CIC-UPES for providing the characterization facility. This work was partially carried out using the facilities of UGC-DAE-CSR Indore. The authors acknowledge the financial support from UGC-DAE-CSR through a Collaborative Research Scheme (CRS) project number CRS/2022-23/01/675, as well as UPES-SEED grant (UPES/R&D-SoAE/25062025/17). We acknowledge DESY (Hamburg, Germany), a member of the Helmholtz Association HGF, for the provision of experimental facilities. Parts of this research were carried out at PETRA III using beamline P22.

**CRediT authorship contribution statement**

**Anmol Sharma:** Writing-review & editing, Writing- original draft, Formal analysis. **Mukul Gupta:** Writing-review & editing, Validation, Resources. **Prasanta Karmakar:** Writing-review & editing, Validation, Investigation, Resources. **Vivek K. Malik:** Writing-review & editing, Validation, Resources. **V. Raghavendra Reddy:** Writing-review & editing, Validation, Investigation, Data curation, Resources. **Andrei Gloskovskii:** Writing-review & editing, Validation, Resources**. Ajay Gupta:** Writing-review & editing, Validation, Methodology, Investigation, Formal analysis, Data curation, Conceptualization. **Ranjeet Kumar Brajpuriya:** Visualization, Validation, Data curation. **Vishakha Kaushik:** Visualization, Validation, Data curation. **Sachin Pathak:** Writing-review & editing, Writing-original draft, Visualization, Validation, Supervision, Methodology, Investigation, Formal analysis, Data curation, Conceptualization.

**Data availability**

Data will be made available on request.